\documentclass[twocolumn,superscriptaddress]{revtex4}
\pdfoutput=1
\usepackage{graphicx}
\usepackage{subfigure}
\usepackage{amsmath}
\usepackage{color}

\begin{document}

\title{Phase diagram of Janus Particles}
\author{Francesco Sciortino} 
\affiliation{ {Dipartimento di Fisica, Universit\`a di Roma {\em La Sapienza}, Piazzale A. Moro 2, 00185 Roma, Italy} }
\author{Achille Giacometti}
\affiliation{Dipartimento di Chimica Fisica, Universit\`a di Venezia, Calle Larga S. Marta DD2137, I-30123 Venezia, Italy}
\author{Giorgio Pastore}
\affiliation{INFM-CNR Democritos and Dipartimento di Fisica dell' Universit\`a,
Strada Costiera 11, 34014 Trieste, Italy}

\begin{abstract}
We deeply investigate a simple model representative of the 
recently synthesized Janus particles, i.e. colloidal spherical particles whose surface is divided  into two areas  of different chemical composition. When the two surfaces are solvophilic and solvophobic, these particles  constitute the simplest example of surfactants.  The phase diagram includes a colloidal-poor (gas) colloidal-rich (liquid) de-mixing region, which is progressively suppressed by the insurgence of micelles, providing the first model where micellization and phase-separation are   simultaneously observed.  {The} coexistence curve is found to be negatively sloped in the temperature-pressure plane, suggesting that  Janus particles can  provide a colloidal system with anomalous thermodynamic behavior.
\end{abstract}


\maketitle

An unprecedented development in particle synthesis is generating a  whole new  set of colloidal particles,   
characterized by   different patterns, particle  patchiness and functionalities\cite{Glotz_Solomon_natmat}.   This extensive effort  is driven  by the  attempt to gain  control over the three-dimensional organization of  self-assembled materials,  characterized by crystalline, or even disordered, colloidal structures  with desired properties.  The recently synthesized Janus 
colloids\cite{Janus,janusgold, janusweitz,Stellacci,granich,janus-softmatter} ---   characterized by a surface divided into two areas  of different chemical composition ---
 offer a fine example of such versatility and  can be reckoned as the paradigm of a system formed by incompatible elements within the same unit structure 
which spontaneously drive a self-assembly aggregation process into complex superstructures. Not only this is ubiquitous in biological
and chemical systems, but, when carried out into a controlled manner, it constitutes one of the most promising bottom-up process for the design of
future materials. Indeed the  
assembly behavior of Janus particles is receiving a considerable attention\cite{cacciuto, phasejanus,cacciuto2}.   

With a proper choice of the  chemico-physical  surface properties, Janus particles can provide the most elementary and geometrically simple example of surfactant particles\cite{erhardt}, in which solvophilic and solvophobic  areas reside on different parts of the surface of the same particle, in a controlled analogy to what is often found in the protein realm, e.g. hydrophobin\cite{hydrophobin} and casein\cite{casein}.

The phase behavior of typical surfactant molecules (as opposed to colloids) has been studied in details in the past, due to their  role in industrial processes and products\cite{bookcolloid}. It is well appreciated that, with the right balance of solvophilic and solvophobic  moieties, surfactants can spontaneously  self assembly into a large variety of structures, including micelles, vesicles, lamellae. Often,
an attractive interaction between the aggregates originates a macroscopic phase
separation into a dense and a dilute micellar phase\cite{corti}.  In rare cases, the surfactant rich phase is characterized by micelles and the surfactant  poor phase by unimers.  One interesting question concerns the relation (or competition) between the gas-liquid phase separation and
the process of {micelles} formation. While it is recognized that {micelles} formation can be seen as a
sort of phase separation (in such a case the liquid is interpreted as an infinite size aggregate)
it is rather unclear how the two phenomena really interrelate and if the same system can show  a clear self assembly process in a certain region of temperatures $T$ and densities  $\rho$ and a  clear gas-liquid phase separation somewhere else.  Previous extensive search for models able to display this mixed behavior--- carried out by Panagiotopoulos and coworkers\cite{thanos1,thanos2,thanos3}  --- has been unsuccessful, {thus} hampering a deep 
 understanding of the interplay between 
aggregation and phase separation. In all studied cases, {on} tuning
some model parameters,   macrophase separation  becomes  replaced by self-organization into isolated or interconnected micelles\cite{likosstar}. 

In this Letter, we show that simply but realistically modeled Janus particles are ideal candidates for
studying, within the same system, the interplay between aggregation and phase separation.
Indeed, the system shows a clear  phase-separation, characterized by the presence of a 
 gas-liquid critical point and a micellar phase at lower temperatures.   Micelles  develop in the gas colloidal phase progressively acquiring  stability,
 suppressing more and more the two-phase coexistence. Differently from all previously studied models,   this {micelles} 
 rich gas phase coexists  with a dense fully connected liquid phase.  
 
   The Janus particles are described via the one-patch model introduced by Kern and Frenkel~\cite{Kern_03}  with coverage $0.5$. In this effective model the two hemispheres of a hard-core particle of diameter $\sigma$ are considered respectively repulsive and attractive. When the segment connecting the  centers of two particles crosses the attractive {hemisphere}  of both particles, {the  two particles then} interact via a square-well potential with range $\Delta=0.5 \sigma$ and depth $u_0$. In all other orientations the interaction is hard-core.
This model  is inspired by  the experimental system investigated in Ref.~\cite{granich}, where the repulsive interaction has an electrostatic origin and the  attractive part is hydrophobic. In the following, $\sigma$ provides the unit of length and $u_0$ the unit of energy. $T$ is also expressed in unit of $u_0$ (i.e. Boltzman constant $k_B=1$).

Results reported in this Letter are based on an extensive numerical study of the
model in a wide range of $T$ and $\rho$.  We employ the most
accurate numerical methodologies to provide a detailed quantification of the phase behavior, including size-effect studies.   Grand-canonical Monte Carlo (GCMC) simulations
(for box size $L=10,12,15,24 \sigma$) have been implemented to locate the critical point and the relation between density and activity,  Gibbs ensemble 
simulations\cite{GEMC} to evaluate the coexisting phases and standard Monte Carlo  (MC) simulations
in the NVT and NPT ensembles   for large system sizes (number of particle $N=5000$) to elucidate the structure of the phases and the equation of state.  
Since 
 self-assembly requires large attraction strengths (compared to $k_BT$) and large systems (indeed aggregates can grow to very large sizes) accurate numerical studies are rather difficult.   
We have explicitly checked that the studied system sizes are  sufficiently large to avoid size effects. Large system sizes  ($N>1000$) have been studied also in GCMC and GEMC, especially at small densities to make sure that the simulation box contains several aggregates
simultaneously. 
In all cases, translational and rotational moves
 consisting of {a} maximum random translation of $\pm 0.1 \sigma$ and
a maximum random rotation of $\pm 0.1$ rad of a randomly selected particle{,} have been implemented. 
Depending on the MC method, insertion and deletion moves   (or swap moves) have been attempted, on average, every 500 
displacement moves  while  volume change moves (with maximum volume changes of the order of $0.5 \sigma^3$) every 100 translational moves. 
Extremely long simulation runs  {(of the order of 10$^8$ MC steps)  have been performed} to reach equilibrium. GEMC and GCMC simulations have been particularly painstaking, due to the  difficulty of inserting/deleting, by a sequence of individual moves, large aggregates, requiring --- 
for each {studied low $T$ state point} --- several months of computer time.

\begin{figure}[ht]
\includegraphics[width=8cm, clip=true]{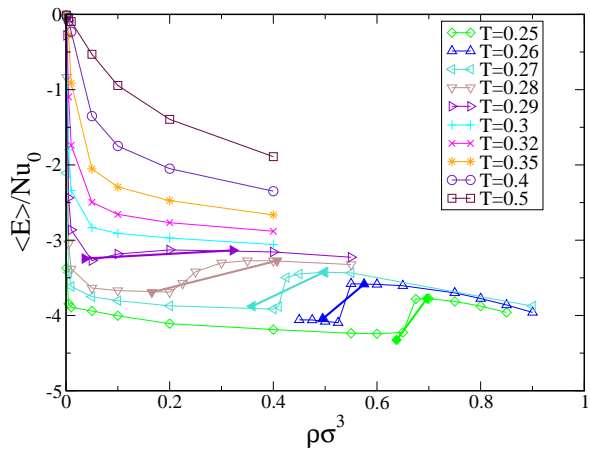}
\caption{Average potential energy $\langle E \rangle$ per particle along several isotherms vs. number  density $\rho \equiv N/L^3$.
Note that  $-2\langle E \rangle/Nu_0 $ coincides with the average number of bonded interactions
per particle.  The thick lines connect the two coexisting densities (full symbols) calculated via GEMC.}
\label{fig:energy}
\end{figure}

\begin{figure*}[hbtp]
\begin{center}
\mbox{
	\leavevmode
	\subfigure [  $\rho \sigma^3=0.001$ ]
	{ \label{f:subfig-1}
	  \includegraphics[width=3.2cm, clip=true]{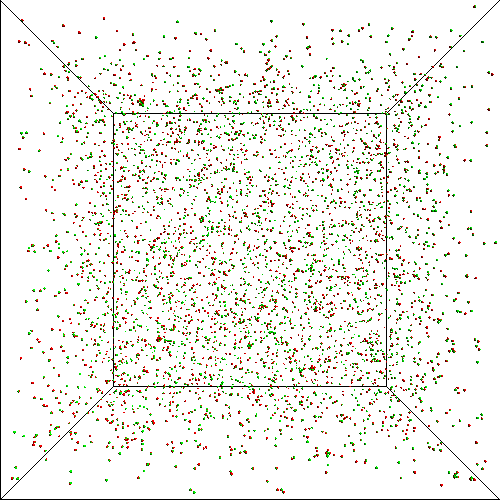} }

		\leavevmode
	\subfigure [  $\rho \sigma^3=0.01$  ]
	{ \label{f:subfig-2}
	  \includegraphics[width=3.2cm, clip=true]{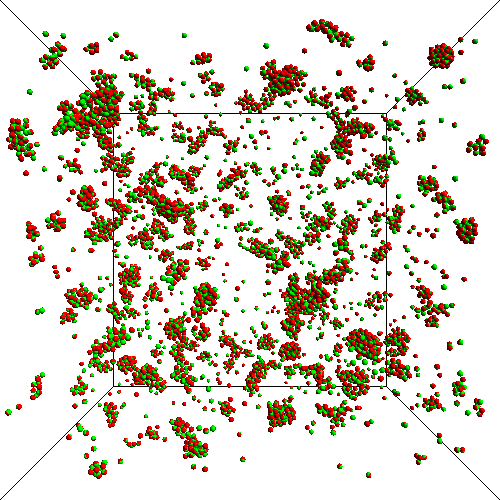} }
	  
	  \leavevmode
	\subfigure [ $\rho \sigma^3=0.1$  ]
	{ \label{f:subfig-3}
	  \includegraphics[width=3.2cm, clip=true]{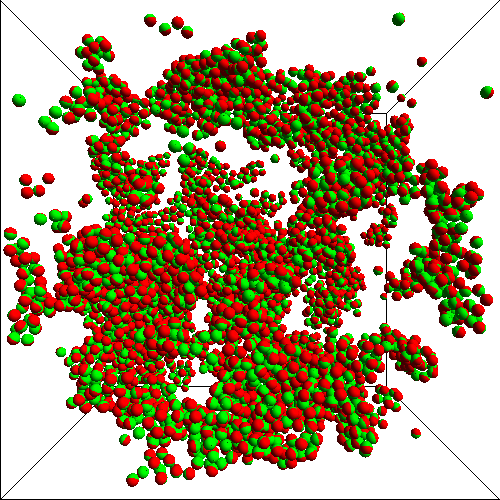} }

\leavevmode
	\subfigure [  $\rho \sigma^3=0.4$  ]
	{ \label{f:subfig-4}
	  \includegraphics[width=3.2cm, clip=true]{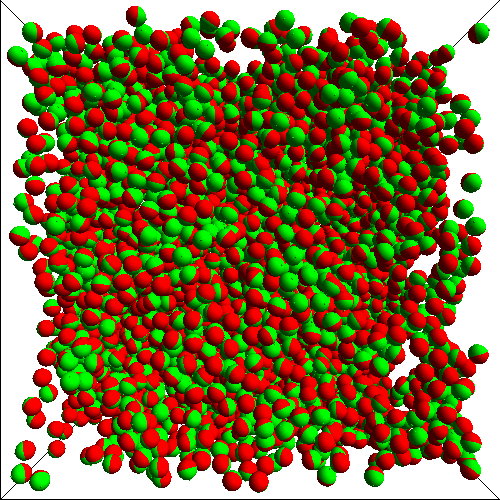} }

\leavevmode
	\subfigure [  $\rho \sigma^3=0.55$ ]
	{ \label{f:subfig-5}
	  \includegraphics[width=3.2cm, clip=true]{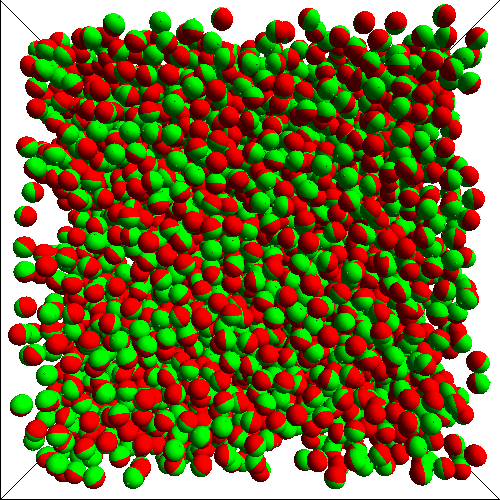} }

}	 
\end{center}
\begin{center}
\mbox{
	\leavevmode
	\subfigure [ $\rho \sigma^3=0.001$ ]
	{ \label{f:subfig-6}
	  \includegraphics[width=3.2cm, clip=true]{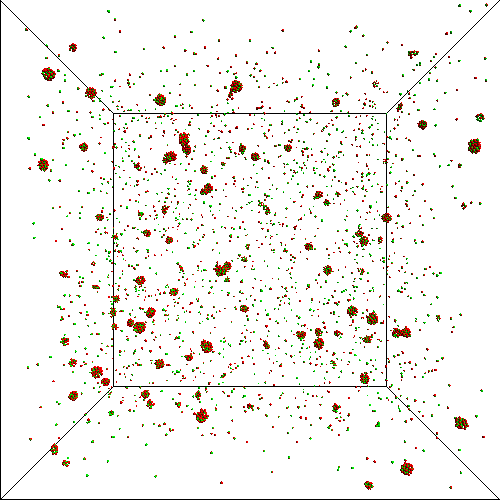} }
	\leavevmode
	\subfigure [  $\rho \sigma^3=0.01$ ]
	{ \label{f:subfig-7}
	  \includegraphics[width=3.2cm, clip=true]{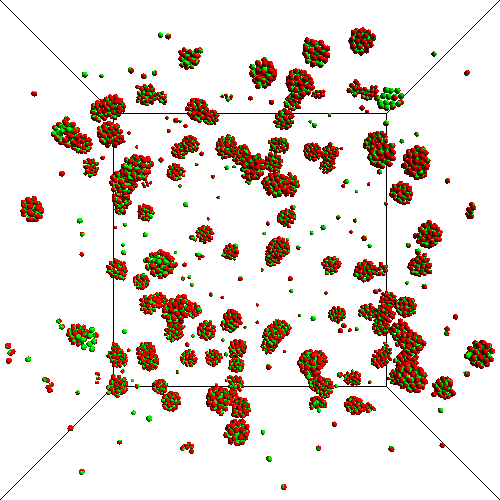} }
	\leavevmode
	\subfigure [  $\rho \sigma^3=0.1$  ]
	{ \label{f:subfig-8}
	  \includegraphics[width=3.2cm, clip=true]{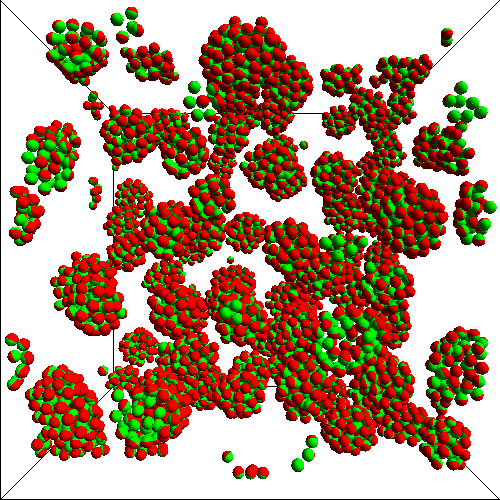} }
	  \leavevmode
	\subfigure [ $\rho \sigma^3=0.4$  ]
	{ \label{f:subfig-9}
	  \includegraphics[width=3.2cm, clip=true]{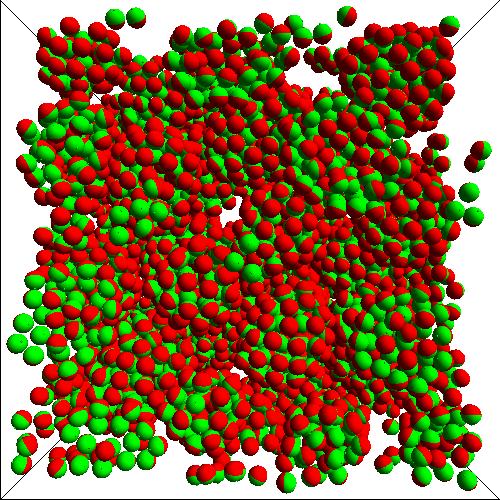} }
       \leavevmode
	\subfigure [  $\rho \sigma^3=0.55$  ]
	{ \label{f:subfig-10}
	  \includegraphics[width=3.2cm, clip=true]{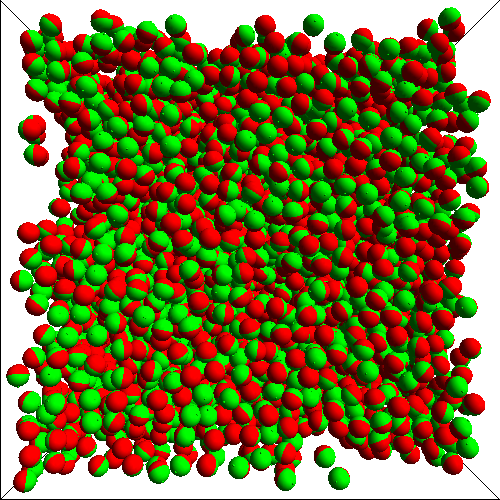} }  

}
\end{center}
\begin{center}
\mbox{
	\leavevmode
	\subfigure [ $\rho \sigma^3=0.001$ ]
	{ \label{f:subfig-11}
	  \includegraphics[width=2.15cm, clip=true]{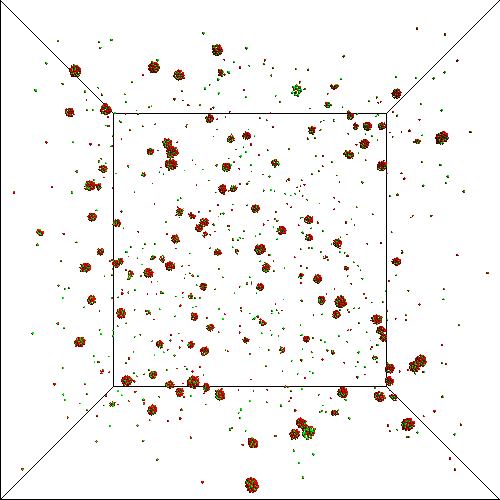} }

	\leavevmode
	\subfigure [  $\rho \sigma^3=0.01$ ]
	{ \label{f:subfig-12}
	  \includegraphics[width=2.15cm, clip=true]{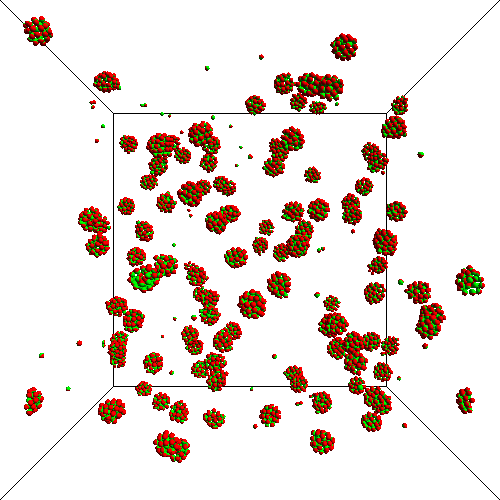} }

	\leavevmode
	\subfigure [  $\rho \sigma^3=0.1$  ]
	{ \label{f:subfig-13}
	  \includegraphics[width=2.15cm, clip=true]{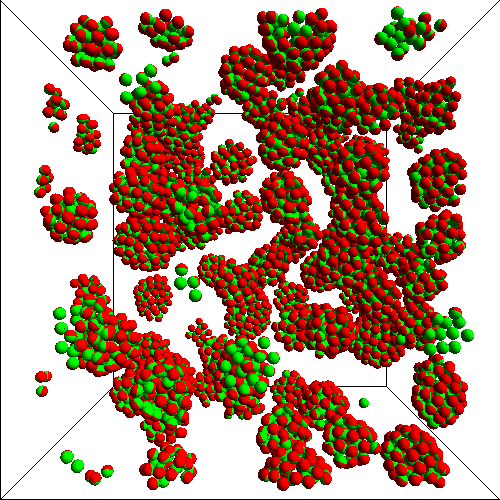} }
	  
	  \leavevmode
	\subfigure [ $\rho \sigma^3=0.4$  ]
	{ \label{f:subfig-14}
	  \includegraphics[width=2.15cm, clip=true]{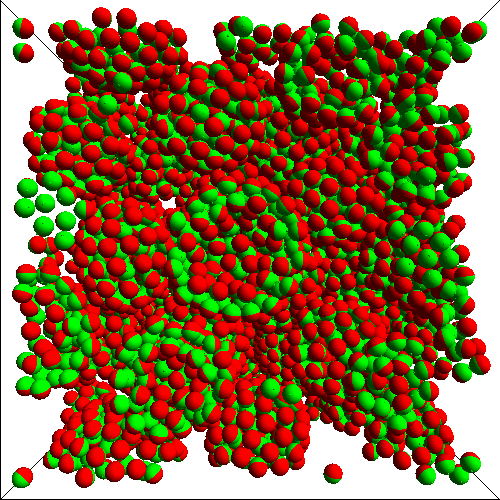} }

	  \leavevmode
	\subfigure [ $\rho \sigma^3=0.55$  ]
	{ \label{f:subfig-15}
	  \includegraphics[width=2.15cm, clip=true]{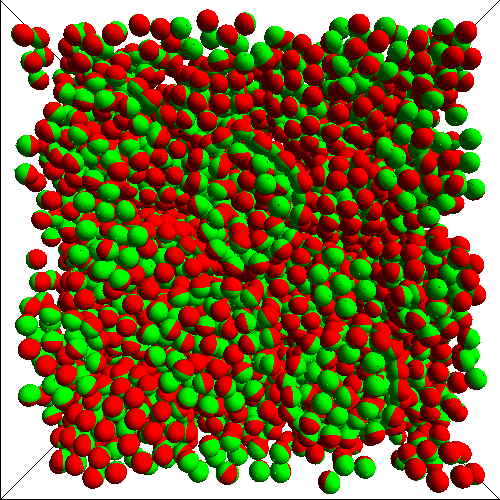} }

      \leavevmode
	\subfigure [ $\rho \sigma^3=0.7$  ]
	{ \label{f:subfig-16}
	  \includegraphics[width=2.15cm, clip=true]{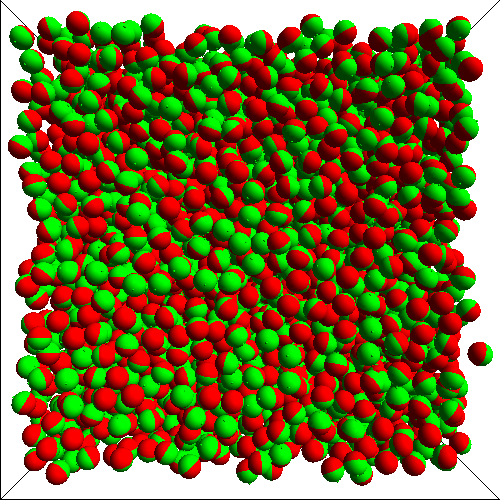} }

\leavevmode
	\subfigure [  $\rho \sigma^3=0.9$  ]
	{ \label{f:subfig-17}
	  \includegraphics[width=2.15cm, clip=true]{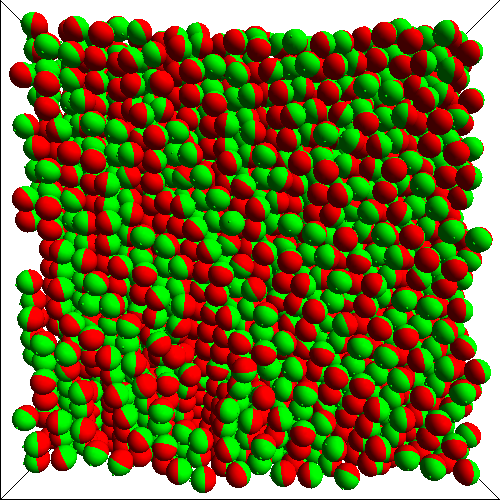} }
} 
\end{center}

\caption{Snapshot of an equilibrium typical configuration for a 5000 particles system at $k_BT/u_0=0.3$ (a-e), $k_BT/u_0=0.27$ (f-j) and  $k_BT/u_0=0.25$ (k-q) for selected densities. The green-coded area is attractive (square-well), the red-coded is repulsive (hard-sphere). The scale changes according to the density.}
\label{fig:snapshot}
\end{figure*}

\begin{figure}[ht]
	     \includegraphics[width=4.cm, clip=true]{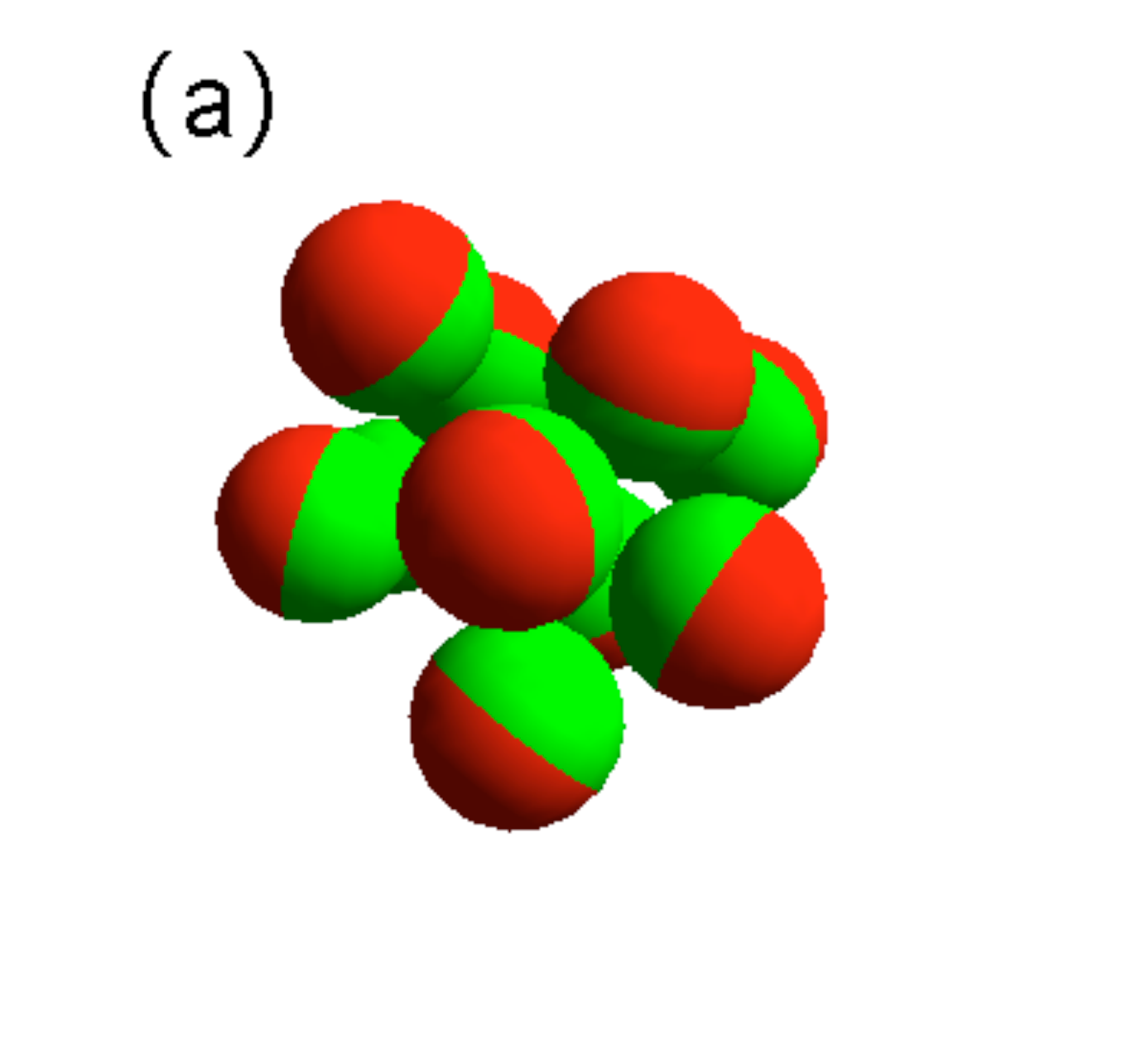} 
	  \includegraphics[width=4.cm, clip=true]{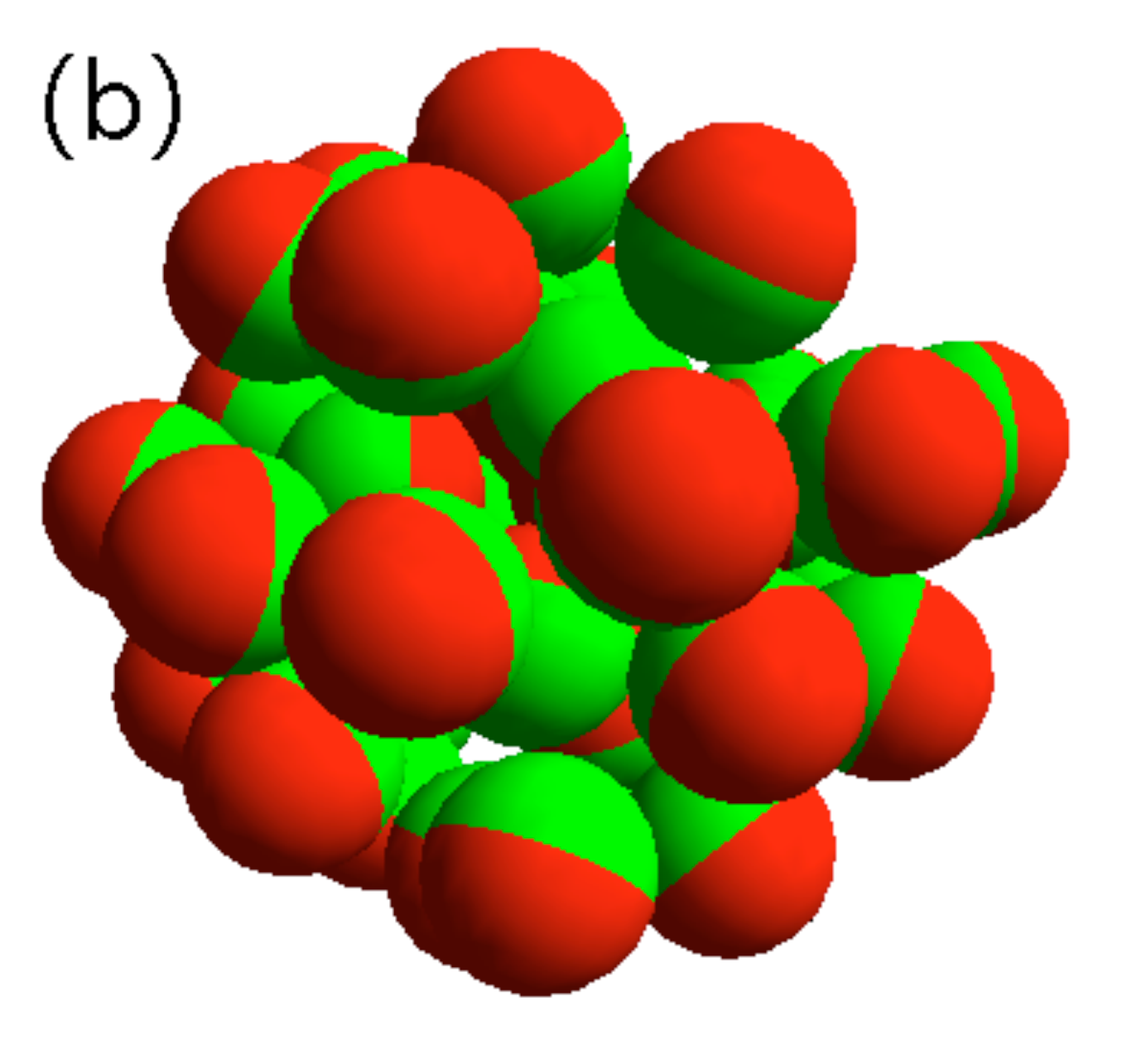} 

\caption{Typical shape of the aggregates of size 10 (a, micelles) and 40 (b, vesicles).
}
\label{fig:shape}
\end{figure}

Fig.~\ref{fig:energy} shows the average potential energy $\langle E \rangle$ per particle along several isotherms.  Below $k_BT/u_0=0.3$, the density dependence of $\langle E \rangle$ becomes non monotonic (reminiscent of the shape of a van der Waals loop), already suggesting the possibility of a phase separation. Several features  significantly differ from the  standard gas-liquid behavior:  the most striking difference is the fact that, close to the loop, particles in the low density states are characterized by $E$ values significantly smaller than the one in the dense phase. At odd with the usual gas-liquid behavior,
the position of the loop shifts to larger $\rho$ and its width  appears to shrink  on  cooling. 
A glance to the structure of the system across density (
Fig.
~\ref{fig:snapshot})  along  different  isotherms
immediately clarifies the reasons for such behavior.  While at {higher} $T$ one observes the  behavior typical of a fluid close to the
gas-liquid critical point, i.e. a spatially inhomogeneous density (or equivalently particles associated in highly poly-disperse clusters), at lower $T$ the system self-assembles
into well-defined{, essentially spherical,} clusters which appear to maintain their 
character in a wide region of $\rho$.  The most frequent clusters are composed by
single-layer micelles  (in which about ten particles orient their attractive part inside the aggregate, 
Fig.~\ref{fig:shape}-(a)
) and double-layer micelles (or vesicles, in which about 40 particles organize into  an inner and an outer shell forming 
a bilayer which exposes to the outside only the hard-core part, preventing 
any attraction between distinct aggregates, Fig.~\ref{fig:shape}-(b)). The double-layer structure becomes the dominant one on cooling and/or on increasing $\rho$.   The organization of the system into these structures 
has the net effect of breaking the thermal correlations which propagate via inter-particle attraction, effectively preventing the development of long range critical fluctuation. 
 This effect is evident at $\rho=0.1$ (Fig.~\ref{fig:snapshot}(c),(h),(m)), where a percolating network of bonded 
 particles is found only at $T=0.30$.  The insurgence of an  energy driven formation of  micelles effectively 
 destabilizes the demixing of the system into a  colloidal-poor (gas) phase coexisting with a  colloidal-rich  (liquid) phase.

\begin{figure}[ht]
\includegraphics[width=8cm, clip=true]{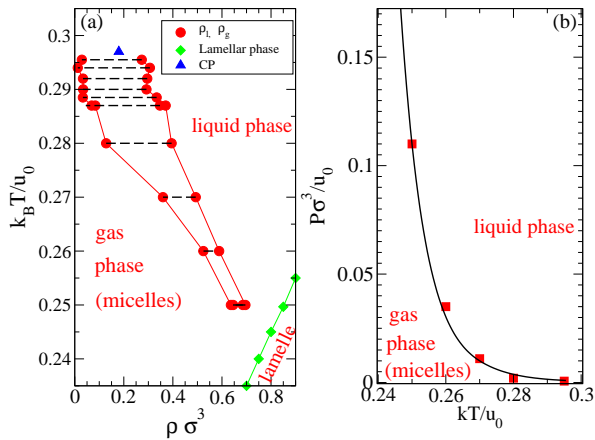}
\caption{Phase diagram  of the Janus model in the $T-\rho$ (a) and $P-T$  (b) planes. Symbols are simulation results, lines are guide to the eyes. }
\label{fig:phase}
\end{figure}

To provide evidence of the existence of a critical point, of  a truly phase-coexistence and to evaluate the appropriate phase boundaries
{,}  we have performed GCMC simulations to 
 locate the $T$ and chemical potential values for which
the density fluctuations behave as the magnetization in the Ising model\cite{Wilding_96} (providing the
location of the critical point) and GEMC simulations, to calculate the {densities} of the
coexisting phases.   Results are reported in  Fig.~\ref{fig:phase}.  
Differently from the standard behavior, the region of phase coexistence which usually widens on cooling,  here shrinks and {shifts} 
to the right, leaving  a wide $\rho$ window   where a homogeneous gas-like phase
composed by  large aggregates [Fig.~\ref{fig:snapshot}~(g)-(h); (m)-(o)]  is thermodynamically stable.  Micelles  coexist with a liquid-like non-micellar phase [Fig.~\ref{fig:snapshot}~(j) and (p)], supporting the evidence that the phase-separation is not related to a critical phenomenon associated with  inter-micelles interactions.  
The densities of both coexisting  phases increase on cooling, progressively approaching each other, suggesting the possibility of a lower $T$ consolute point.  Unfortunately,
both the gas of micelles and the liquid phase become metastable in respect to 
the formation of a ordered (crystalline, lamellar) phase; the onset of crystallization, {on} the time scale probed by the numerical calculations, pre-empts the possibility of exploring such possibility. A snapshot of the system ordered in the  lamellar phase is reported in 
Fig.~\ref{fig:snapshot}(q). 

\begin{figure}[ht]
\includegraphics[width=8cm, clip=true]{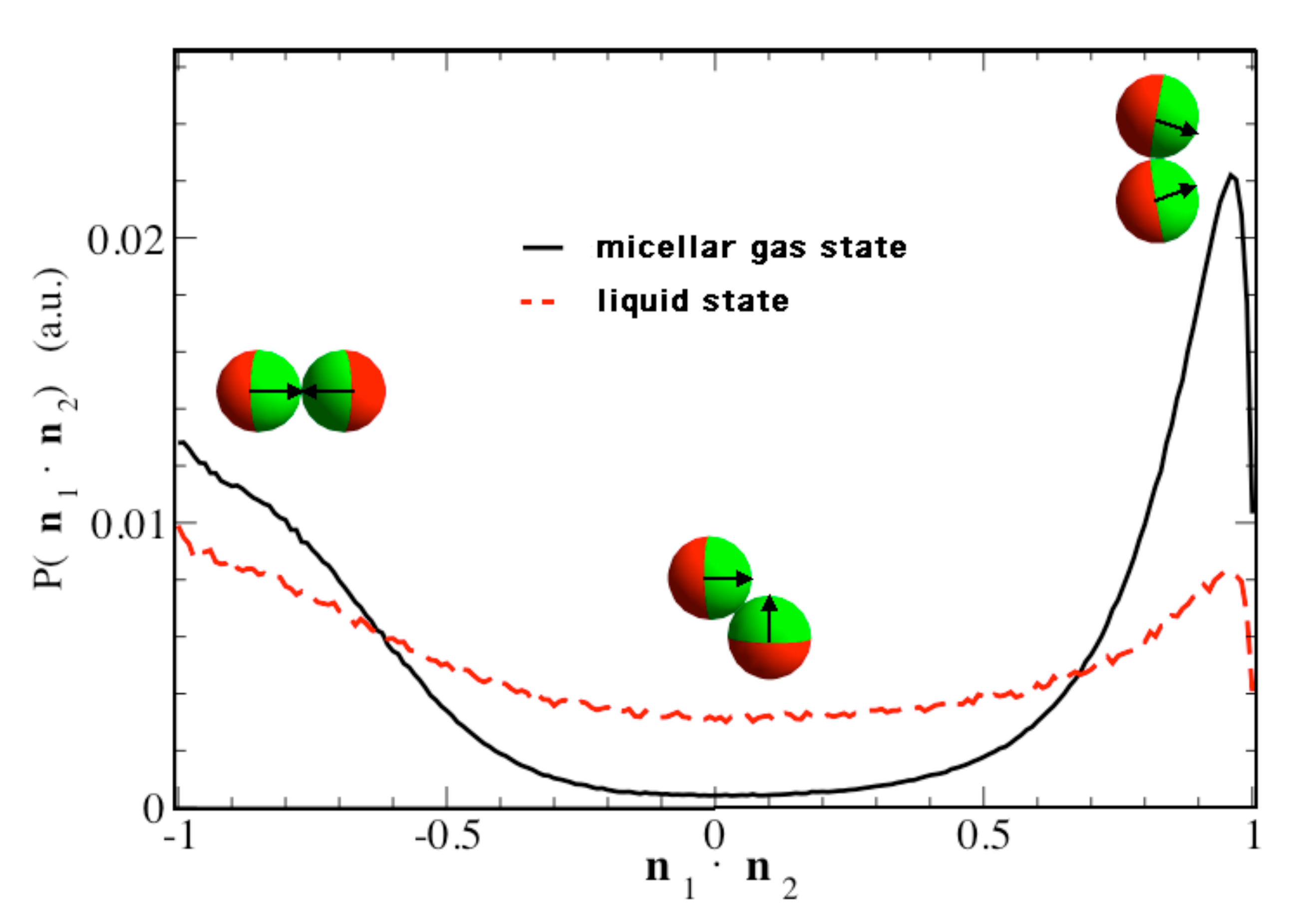}
\caption{Distribution of  the scalar product ${\bf n}_1 \cdot {\bf n}_2$ evaluated over all pairs of bonded particles in a gas ($\rho\sigma^3=0.65$) and in a liquid   ($\rho\sigma^3=0.675$) state point, both at $kT/u_0=0.25$. Cartoons provide a sketch of the relative orientations of the bonded particles. Arrows  indicate the corresponding {unit vectors}.}
\label{fig:order}
\end{figure}

The energetic and entropic balance associated with the transition from the 
micelles gas to the liquid phase reveals some important information.  As previously alluded while discussing Fig.~\ref{fig:energy},  differently from the
usual gas-liquid behavior,  $\langle E \rangle$ is higher in the liquid phase than in the
micellar gas phase.  Hence, despite the fact that the gas phase is stabilized by the translational entropy of the micelles, at coexistence the liquid phase is probably more  (orientational) disordered than the gas one. Insights on the larger orientational order in the
gas phase as compared to the liquid one, can be obtained by computing the
distribution of relative orientations between all pairs of bonded particles. More precisely, for all bonded pairs, we have evaluated the distribution of the scalar product ${\bf n}_1 \cdot {\bf n}_2$ where ${\bf n}_1$ and ${\bf n}_2$  are the two {unit vectors} indicating the location of the patch center in each particle frame (see arrows in Fig.~\ref{fig:order}). 
The distribution $P({\bf n_1} \cdot {\bf n_2})$ is expected to show well defined peaks
for an ordered state and to be flat in a completely disordered  state.
As shown in Fig.~\ref{fig:order},  $P({\bf n_1} \cdot {\bf n_2})$ is indeed significantly more structured in the  micelles-rich gas phase 
with respect to the liquid phase.
 
To provide a definitive confirmation of the lower entropy of the gas phase ($S_g$)  as compared to the liquid  one ($S_l$), we evaluated the equation of state at several $T$ by mean of MC simulations in the PNT ensemble. These results, coupled with the
 coexisting densities evaluated via GEMC, provide a method to 
 calculate the slope ($dP/dT|_x$) of the coexistence curve  in the $P-T$ plane.
By virtue of the Clapeyron equation, $dP/dT|_x$  is a measure of $\Delta S/\Delta V$, where  $\Delta S$ and $\Delta V$ are the differences in entropy and volume of the two coexisting phases. Since $V_g>V_l$ a positive (negative) slope implies that $S_g > (<) S_l$.  
Fig.~\ref{fig:phase}-(b) shows that the  coexistence curve in the $(T-P)$ plane has an anomalous behavior. Differently from the usual gas-liquid coexistence, here the curve is negatively sloped, confirming that $S_g<S_l$, despite the significantly larger translational contribution.  
 The explanation for this odd behavior can be traced back to the orientational ordering of the particles in the gas micellar phase  and to the  disordered orientations which characterize the liquid phase (Fig.~\ref{fig:order}). As a result, the orientational disorder gain in the fluid phase  compensates the energetic penalty. The negatively sloped gas-liquid coexistence curve indicates also  that, on cooling along a constant (osmotic) pressure path, the density of the system decreases at the transition.

The results presented here have far-reaching consequences in many respects. From the theoretical side,  our work constitutes the first study
displaying a clear gas-liquid critical point which evolves into phase coexistence between a cluster and a liquid phase through a competition between
phase-separation and micelles formation. In addition, the system studied here behaves as  
anomalous substance in which the dense phase is more disordered 
than the low-density phase and which,  in analogy with the negatively sloped liquid water-hexagonal ice coexistence curve, expands  on cooling along isobars. 
The simplicity of the model and its rich and unconventional phase diagram {begs for experimental verifications}.  Despite the investigated model has a range  ($0.5 \sigma$) larger than the  interaction ranges  typical of  $\mu$m-sized  colloids (usually of the order of 0.05-0.2 $\sigma$), ranges comparable to $0.5 \sigma$ can be realized with nanoparticles or building on  the recently explored Casimir critical forces~\cite{casimir1,casimir2}.

We acknowledge support from NoE SoftComp NMP3-CT-2004-502235, ERC-226207-PATCHYCOLLOIDS and PRIN-COFIN 2007. We thank C. De Michele  for providing us with the code for generating  Fig.~\ref{fig:snapshot} and  J. Horbach,  J. Largo and M. Noro for discussions.

\end{document}